\pdfoutput=1 
\documentclass[12pt]{article}

\usepackage{amsmath}
\usepackage{amssymb}
\usepackage{euscript}

\usepackage{inputenc}

\usepackage{amsmath,amssymb,bbm,color}
\usepackage[T1]{fontenc}
\usepackage{amsbsy}
\usepackage{euscript}
\usepackage{graphicx}
\usepackage{hyperref}
\usepackage{cite}
\usepackage{enumerate}
\numberwithin{equation}{section}

\begin{document}
\vsize=25.0 true cm
\hsize=16.0 true cm
\predisplaypenalty=0
\abovedisplayskip=3mm plus 6pt minus 4pt
\belowdisplayskip=3mm plus 6pt minus 4pt
\abovedisplayshortskip=0mm plus 6pt
\belowdisplayshortskip=2mm plus 6pt minus 4pt
\normalbaselineskip=14pt
\normalbaselines
\begin{titlepage}

\vspace{-4.0cm}
\begin{flushright}
		{\bf IFJPAN-IV-2020-05}   \\
                {\bf BU-HEPP-2020-01}
\end{flushright}
\vspace{5mm}
 
\vspace{5mm}
\begin{center}
{\bf\large
The Monte Carlo Program {\tt KKMC},\\
for the Lepton or Quark Pair Production at LEP/SLC Energies -- \\
updates of electroweak calculations
}\end{center}

 \vspace{1mm}
\begin{center}
{       A. Arbuzov$^{a}$,
        S.\ Jadach$^b$,
        Z.\ W\c{a}s$^b$,
        B.F.L.\ Ward$^{c}$ and 
        S.A.\ Yost,$^d$ }
\\
\vspace{2mm}
{\em $^a$Joint Institute for Nuclear Research,
Joliot-Curie str. 6, 141980, Dubna, Russia}
\\
{\em $^b$Institute of Nuclear Physics, Polish Academy of Sciences,\\
  ul.\ Radzikowskiego 152, 31-342 Krak\'ow, Poland}
\\
\vspace{1mm}
{\em $^c$Department of Physics,
   Baylor University, Waco, TX 76798-7316, USA}
\\
\vspace{1mm}
{\em $^d$The Citadel, Charleston, SC, USA}
\end{center} 
\vspace{5mm}
\begin{center}
{\bf   ABSTRACT}
\end{center}
Since the {\tt KKMC} program was published for the first time over 20 years ago,
it has gained popularity and was exploited in a broad spectrum of applications. 
The core part of the program itself did not change much. 
In contrast, some of the libraries have evolved substantially.

The aim of this publication is to archive four versions, 
alternative to the one published 20 years ago
versions of the electroweak libraries (or just parameter initialization versions),
which were instrumental for the precision Standard Model calculation
from the end of LEP era till now and
for the sake of the future applications/comparisons for
the future electron-positron colliders,
in particular for the  FCC-ee related studies.
These electroweak libraries are useful for the hadron collider applications as well, 
for instance for {\tt KKMC-hh} or {\tt TauSpinner} projects.

\vspace{6mm}
\centerline{\it To be submitted to Computer  Physics Communications}

\vspace{8mm}
\begin{flushleft}
{\bf 
 IFJPAN-IV-2020-05, BU-HEPP-20-01\\
 May 2020}
\end{flushleft}

\footnoterule
\noindent
{\footnotesize
\begin{itemize}
\item[${\dag}$]
This work is partly supported by
 the Polish National Science Center grant 2016/23/B/ST2/03927,
  under decisions DEC-2017/27/B/ST2/01391
 and the CERN FCC Design Study Programme. 
\end{itemize}
}

\end{titlepage}

\noindent{\bf UPDATE SUMMARY}
\hfill\vskip 10pt
\noindent{\sl Title of the program:} {\tt KKMC}\ .
 
\noindent{\sl Reference to original program:}
Comput. Phys. Commun. {\bf 130} (2000) 260
 
\noindent{\sl Authors of original program:}
              S. Jadach, B.F.L. Ward and Z. W\c as
 
 
\noindent{\sl Operating system:}
LINUX
 
\noindent{\sl Programming language used:}
FORTRAN 77
 
\noindent{\sl High-speed storage required:}  $<$ 1 MB
 
\noindent{\sl No. of bits in a word:} 32, 64
 
\noindent{\sl Peripherals used:} Line printer
 
 
 
\noindent{\sl Nature of the physical problem:}
Spin polarization of the $\tau$ in the process
$e^+e^-\rightarrow $ $\tau^+\tau^-(n \gamma ),$
                      $\tau^\pm\rightarrow X^\pm$
is used as an important data point for precise tests
of the standard electroweak theory. 
The  effects due to QED bremsstrahlung and apparatus efficiency have
to be subtracted from the data.
The program applies, as well, to the muon and neutrino pair production processes.
It can simulate also the electron positron annihilation into  
$u$, $d$, $s$, $c$, $b$ quark pairs.
An important segment of the calculations rely on electroweak loop effects
provided by the appropriate program library.

\noindent{\sl Method of solution:}
The Monte Carlo simulation of the combined $\tau$ production and decay
process is used to calculate the spin effects and effects of radiative
corrections, including hard bremsstrahlung, simultaneously.
Any experimental cut  and apparatus efficiency may easily be introduced
by rejecting some  of the generated events. Electroweak effects are provided
with an external library with the help of disk stored lookup tables.
 
\noindent{\sl Restrictions on the complexity of the problem:}
The high precision of the program is assured
in the region near the $Z$ resonance.
For other energy ranges it varies. 
At low energies predictions from electroweak libraries need 
to be replaced with an appropriate dedicated code.  \\
We document variants of electroweak libraries introduced over the  years since
{\tt KKMC} was published [1] leaving other changes to further publications.
The distribution package provided now corresponds to KKMC version 4.16
and its physics content is documented in Ref. [2].
Details other than the electroweak details of
this version are not addressed, 
they do not differ much from those of Ref.~[1].

\vspace{1mm}
\noindent
[1] S.~Jadach, B.F.L.~Ward and Z.~Was,
Comput. Phys. Commun. \textbf{130} (2000), 260-325

\noindent
[2] S.~Jadach, B.F.L.~Ward and Z.~Was,
Phys. Rev. D \textbf{63} (2001), 113009

\newpage

The {\tt KKMC} Monte Carlo \cite{Jadach:1999vf} is designed to simulate two fermion
production process in the electron-positron colliders,
$e^+e^- \to 2f\ n\gamma$, ($f=\mu,\tau,u,d,c,s,b, \nu,\; f\neq e$ ).
It is armed with the most advanced QED matrix element
based on the coherent exclusive exponentiation (CEEX)
of  the initial and final state bremsstrahlung,
valid up to the highest FCC-ee energies \cite{Jadach:1998jb,Jadach:2000ir}.

The detailed description of how to use
the  {\tt KKMC} Monte Carlo program can be found in Ref.~\cite{Jadach:1999vf},
The physics content of the version 4.16 which we will use in the present paper 
as a reference is explained in Ref.~\cite{Jadach:2000ir}.
For further details  see references therein 
and  {\tt KKMC}  webpage \cite{jadach-www},
where  up to date envelopment versions of {\tt KKMC}  can also be  found. 
For novel applications and for further references see e.g. Ref.~\cite{Ward:2018qkh}.

The differences of the new {\tt KKMC} program versions, 
with respect to the published ones, are not big --
the structure of the program and user interface have not changed.
In particular the methodology of interfacing
electroweak corrections calculation to the core {\tt KKMC} code
still follows the prescription given in Ref.~\cite{Jadach:1999vf}.

Now, when the FCC effort takes momentum, 
it is a good time to archive the  {\tt KKMC} program,
its cross-checks and documentation, for the future references.
The important first step in this direction is the proper archiving
of the variants of electroweak libraries  
used in {\tt KKMC} program
over the past two decades and at present.
Let us keep in mind that the recently developed offspring 
program {\tt KKMC-hh}  \cite{Jadach:2016zsp} 
for Z production in hadron collider applications 
also uses the same electroweak (EW) libraries.
In addition, {\tt TauSpinner} of ref.~\cite{Richter-Was:2018lld}
exploits EW results obtained from the EW libraries update presented here.
The present paper may be treated as an appendix to  \cite{Jadach:1999vf} 
rather than as an independent publication.

Already nowadays, work on experimental  tests of Standard Model in ATLAS
turned out to pose some challenges for proper
adjustment of electroweak software and to conventions of LEP 1 and LEP 2 times
\cite{ATLAS-CONF-2018-037}. 
This experience contributes also to our motivation.
This may be even more important in the future, when expertise of people
involved in LEP efforts will be less available than now.

In the following, a minimal information on the content of the
upgrade will be provided.

\vspace{2mm}
\centerline{\bf Electroweak libraries}
\vspace{2mm}

The  electroweak library {\tt DIZET} version 6.21  
\cite{Bardin:1989tq,Bardin:1999yd} was installed in 
the  {\tt KKMC} from the very beginning
in versions of Refs.~\cite{Jadach:1999vf,Jadach:2000ir}.
Since then the {\tt KKMC} code includes $ WW $  and $ ZZ $ boxes and other non-QED
corrections such as top loop/vertex corrections important for precision
predictions. 
The basics of the original  {\tt DIZET} interface did not require modifications
and is the same until the present days%
\footnote{Note however that
the interface of the {\tt DIZET} library to the function calculating the contribution of
low energy $e^+e^- \to hadrons$ data varies between version 6.21 and later
versions.}.

Luckily, the pretabulation procedure (interface) of the electroweak form factors,
that is in the form of the lookup tables in the disk files,
used in {\tt KKMC} offers an easy way for the upgrades 
with the newer versions of the {\tt DIZET} library. 
In the {\tt KKMC} program these lookup tables
can be also optionally produced in flight,
instead of being stored in text disk files.
However, for the sake of archivization, 
the version with lookup tables on the disk is included in the present distribution
because it demonstrates manifestly how well the EW library is independent 
from the rest of the MC code and also has some practical advantages --
the  electroweak initialization
can be more easily adjusted locally for each electroweak library variant.

Using the pretabulating algorithm for the {\tt DIZET} version 6.21
of Ref.~\cite{Jadach:1999vf} the presented distribution package includes
{\tt KKMC} version 4.16d  compatible with Ref.~\cite{Jadach:1999vf}
and several versions of the {\tt DIZET} library:
\begin{enumerate}
\item
Version 6.21 with updated input parameters, 
directory {\tt dizet-6.21}.
A thorough verification of the implementation
{\tt DIZET} 6.21 into KKMC was
performed in Ref.~\cite{Kobel:2000aw}.
\item
Version 6.42~\cite{Arbuzov:2005ma} used at
the time when final LEP data was analyzed, directory {\tt dizet-6.42-cpc}.
This code was published with the 
version of hadronic contribution to virtual photon
vacuum polarization \cite{Eidelman:1995ny},
obsolete already at that publication time.
However, its importance is that it is the last published version of the code
and could be useful to reproduce 
some old published benchmarks \cite{Jadach:1999vf,Kobel:2000aw}.
\item
Version 6.42 with the updated vacuum polarization of Ref.~\cite{Jegerlehner:2017zsb},
directory {\tt dizet-6.42}.
\item
Version 6.45 of Ref.~\cite{dz6-45}, directory {\tt dizet-6.45}.
\end{enumerate}
It should be stressed that
each variant of the {\tt DIZET} library includes a specific variant 
for the  {\tt dizet-xxx/input.data} file,
which redefines a few default input parameters
defined in the {\tt .KK2f\_defaults} file%
\footnote{%
In case of the use of predefined EW lookup tables, these redefinitions
have to be repeated one more time in the user input file of {\tt KKMC}
used for the Monte Carlo generation run.}.
Version specific parameters in the {\tt .KK2f\_defaults} of {\tt KKMC}
are for version 6.21,
but constants like masses the of $Z$ boson, Higgs boson and top quark,
masses of other fermions and
the QCD coupling constant are already updated to the present PDG values%
\footnote{The original version is kept for the record as {\tt .KK2f\_defaults-2000}.}.

At the lower energies, e.g. those of Belle-II it is required that
the EW calculations are replaced by the fine tuned prediction for
the photon vacuum polarization.
Related  issues are covered in the work of Ref.\cite{Banerjee:2007is},
being the most up to date version of the public archivization.
It is not integrated into presented upgrade,
as this is mainly targeting the needs of future high energy projects,
while Belle-II is an on-going project.

Note that the {\tt SANC} project~\cite{Andonov:2004hi,sanc:2020} 
is now the vigorous continuation
of the {\tt DIZET} and {\tt ZFITTER} project.

\vspace{2mm}
\centerline{\bf How to install and run}
\vspace{2mm}
The  main directory ({\tt Dizet\_Upgrades\_in\_KKMC4.16)} in the distribution tarball 
includes the {\tt KKMC-v.4.16e} directory and next to it the directories 
{\tt dizet-6.21}, {\tt dizet-6.42}, {\tt dizet-6.42-cpc} and {\tt dizet-6.45}.

The {\tt KKMC} program version 4.16 does not differ much from
the version 4.13 published in Ref.~\cite{Jadach:1999vf}.
Version 4.13 is the one that was used during 
the LEP workshop 1999/2000, see Ref.\cite{Kobel:2000aw}.
It corresponds also closely to documentation of
the physics content of {\tt KKMC} in Ref.~\cite{Jadach:2000ir}
and to Ref.~\cite{Jadach:2000ir}.

The {\tt KKMC-v.4.16e} directory does not include subdirectory {\tt dizet}.
It is now placed outside and renamed as {\tt dizet-6.21}.
It is the original EW library {\tt DIZET} version 6.21 published in
Refs.~\cite{Bardin:1989tq,Bardin:1999yd} 
and present in the code of Ref.~\cite{Jadach:1999vf}.
(For input parameter initialization see below).
The other three newer versions of {\tt DIZET},
{\tt dizet-6.42}, {\tt dizet-6.42-cpc} and {\tt dizet-6.45}.
are also placed not directly inside the {\tt KKMC-v.4.16e} directory
but outside, next to it.

The source codes of all four {\tt DIZET} directories
do not require any new documentation, because
the description given in \cite{Jadach:1999vf} for interfacing the EW correction
into the matrix element of {\tt KKMC} and of the methodology 
of the use of the lookup tables of the EW formfactors in {\tt KKMC}
remains valid for all the above new EW directories.

The appropriate interface subprograms creating EW lookup tables
are executed inside each of the above listed directories of the choice.
The interface subprograms {\tt TabMain.f} and {\tt DZface.f}
are compiled and executed independently of the main {\tt KKMC} program.
The only connection with the main {\tt KKMC} source code
is that they read physics input parameters and other configuration parameters
from the default input data of {\tt KKMC}
encoded in the file {\tt .KK2f\_defaults}.
The header file {\tt BornV.h} which defines the range and density
of the lookup tables must be identical 
in the interface programs creating tables in a given {\tt DIZET} directory
and in the {\tt bornv} subdirectory of the {\tt KKMC-v.4.16e} directory,
otherwise {\tt KKMC} will stop and print a message.

Switching form one EW library to another is described below.
For a given EW library one should create EW tables locally by means 
of executing the {\tt make  table.all} command 
in the directory of the relevant EW library.

Let us explain carefully the structure of the input parameters.
First of all, {\tt TabMain.f} reads default data and parameters
from the {\tt KKMC-v.4.16e/.KK2f\_defaults} file of the {\tt KKMC}.
The values of the physics constants like masses of particles are
updated in this file to present PDG values.
In case the user would like to 
check backward compatibility with old benchmarks
we also keep the original version as {\tt .KK2f\_defaults-2000}.
Next {\tt TabMain.f} reads the local {\tt dizet-x.yy/input.all} file
and input data in this file overwrites the default values taken from
{\tt .KK2f\_defaults}. 
The main purpose of {\tt dizet-x.yy/input.all} is to adjust
steering parameters specific for a given version of {\tt DIZET}
and following the preferences of the user. 
Tables on the disk are created with the
input data from global {\tt .KK2f\_defaults} and local {\tt input.all}
take preference.

Once EW look-up tables are created, 
then a simple benchmark run of {\tt KKMC} in 
\\
the {\tt KKMC-v.4.16e/ffbench/} can be executed
following instructions in 
\\
the {\tt KKMC-v.4.16e/ffbench/HowToStart} file.
Note that in the Monte Carlo run using lookup tables
{\tt KKMC} will read one more time the {\tt .KK2f\_defaults} file and
possibly some user input data in the work directory of the MC run.
It is the user's responsibility to take care that the above user input data 
of the MC run are compatible, or even identical, 
with input data used during the creation of the lookup EW tables%
\footnote{
{\tt KKMC} will try to detect mismatch of the input data
in pretabulation and in the MC run,
but this is not a completely foolproof procedure.
}.

In order to facilitate switching quickly from one EW library to another,
we have automatized it adding extra functionality to 
{\tt KKMC-v.4.16e/ffbench/Makefile}.
For example staying in the {\tt ffbench} directory
while executing the command {\tt make link-dizet-6.45}
creates convenient {\em soft links} to {\tt ../dizet-6.45} directory
and executing next {\tt make EWtables} will create new EW lookup tables.
Similar commands like {\tt make link-dizet-6.21} are available.
In the above scenario switching from one to another EW library is fast and easy.
Summarizing,
after unpacking {\tt Dizet\_Upgrades\_for\_KKMC4.16.tgz},
the interested user may execute the following set of commands
in order to check the integrity of the provided source code:
\begin{center}
{\tt
\begin{tabular}{  l  l l}
 cd   & KKMC-v.4.16e/ffbench/ \\
 make & link-dizet-6.45 & ~~~{\rm creating soft links}\\
 make & makflag         & ~~~{\rm updating compiler flags everywhere}\\
 make & EWtables        & ~~~{\rm creating EW lookup tables}\\
 make & demo-start      & ~~~{\rm running short MC run}
\end{tabular}
}
\end{center}
In case of switching to another library, 
for example  with {\tt make link-dizet-6.21}, 
one should do {\tt make Clean}
and repeat the above sequence of commands one more time.

Finally, there is also an available option in {\tt KKMC}, 
of the {\em in flight initialization} of the EW lookup tables,
placing them directly in the fortran common blocks,
without storing them in the disk files%
\footnote{The in flight creation of EW tables (without 'make EWtable')
for the use in {\tt KKMC} can be activated very easily,
by means of uncommenting {\tt EXT\_LIBRARYd} 
in the ffbench/Makefile.}.
In this case input data and parameters in {\tt input.all} are ignored,
hence the user may not worry about possible mismatch between input
used during table creation phase and their use in the MC generation run
(or some other application using them).
All input parameters are taken from {\tt .KK2f\_defaults} and 
are corrected/updated by the user input of the actual MC run,
which obviously must fit the type of the EW library actually being used.
This option is more convenient for the {\tt KKMC} user who is using all the time
just one EW library.
The in flight creation of EW tables takes only a few seconds of CPU time.

\vspace{5mm}
\centerline{\bf\large Acknowledgments}
\vspace{2mm}
We acknowledge work of E. Richter-Was which was important
for the completion of the  present work.
We would like to thank all authors of libraries used in the program for their
cooperation and support, especially the authors of {\tt DIZET}.
This work is partly supported by
the Polish National Science Center grant 2016/23/B/ST2/03927,
under decisions DEC-2017/27/B/ST2/01391
and the CERN FCC Design Study Programme.
A. Arbuzov is grateful for support to the RFBR grant 20-02-00441.


\end{document}